# Millimetre-waves to Terahertz SISO and MIMO Continuous Variable Quantum Key Distribution

Mingqi Zhang[1], Stefano Pirandola[2], and Kaveh Delfanazari[1,*]

[1] *Electronics and Nanoscale Engineering Division, James Watt School of Engineering, University of Glasgow, Glasgow G12 8QQ, UK*
[2] *Department of Computer Science, University of York, York YO10 5GH, UK*
*Corresponding author: kaveh.delfanazari@glasgow.ac.uk
Dated:25122022

*Abstract*—With the exponentially increased demands for large bandwidth, it is important to think about the best network platform as well as the security and privacy of the information in communication networks. Millimetre (mm)-waves and terahertz (THz) with high carrier frequency are proposed as the enabling technologies to overcome Shannon's channel capacity limit of existing communication systems by providing ultrawide bandwidth signals. Mm-waves and THz are also able to build wireless links compatible with optical communication systems. However, most solid-state components that can operate reasonably efficiently at these frequency ranges (100GHz-10THz), especially sources and detectors, require cryogenic cooling, as is a requirement for most quantum systems. Here, we show that secure mm-waves and THz QKD can be achieved when the sources and detectors operate at cryogenic temperatures down to $T$= 4K. We compare single-input single-output (SISO) and multiple-input multiple-output (MIMO) Continuous Variable THz Quantum Key Distribution (CVQKD) schemes and find the positive secret key rate in the frequency ranges between $f$=100 GHz and 1 THz. Moreover, we find that the maximum transmission distance could be extended, the secret key rate could be improved in lower temperatures, and achieve a maximum secrete communication distance of more than 5Km at $f$=100GHz and $T$=4K by using 1024×1024 antennas. Our results may contribute to the efforts to develop next-generation secure wireless communication systems and quantum internet for applications from inter-satellite and deep space, to indoor and short-distance communications.

## I. INTRODUCTION

With the extension of wireless communication and the fast development of information security, higher carrier frequencies and more spectral resources are required [1, 2]. Millimetre (mm)- and terahertz (THz)- waves [3-6] offer ultrawide bandwidth and high-speed data rate communication and are considered to build next-generation (6G) communication systems [7-11]. Mm-waves and THz bands lie between the mature microwave and optical bands as less explored area [12-16]. A gap in the electromagnetic spectrum exists at these frequency ranges due to the inefficient and unpractical of the devices and circuit [1-19]. However, the recent development of electronic, photonic and plasmonic-based mm-waves and THz technologies help close this gap with the demonstration of power-efficient sources [20-26], antennas [27-31], filters [32-34], waveguides [29, 35-39], modulators [40-49], and detectors [3, 49-52]. Demands for 6G are including, but are not limited to, Terabit per second (Tb/s), mm-precision sensing and positioning, seamless connectivity, and ultrafast wireless communications [7-11]. Moreover, practical implementation of quantum processors and quantum computers operating at low temperatures (cryogenics) [53-55], requires massive open air and free space data transfer from and to high-performance classical processors, computers, and communication systems. Therefore, to realise a robust building block for practical quantum information processing attention should be on both security and low-temperature operation. Compared with the free-space optical link, the THz link is more stable under harsh environments such as fog conditions [56]. The limit of mm-waves [57] and THz links [58, 59] in long distances is mainly caused by the absorption of the air [60]. So it is important to find the window with low atmospheric absorption through this band. High-level security is also an important aspect of realising mm-waves and THz communications which is quite challenging to maintain with classical cryptography schemes. Quantum key distribution (QKD) can help to achieve the goal of high-level unconditional security with the power of the quantum physics [61-64]. QKD could be divided into discrete variables (DVQKD based on single photon sources and detectors) and continuous variables (CVQKD based on standard communication systems) [62-75]. CVQKD uses coherent homodyne detection instead of single photon detection [65] and could be integrated with next-generation communication systems [66].

Single-input single-output (SISO) is a kind of classical communication system where the transmitter and receiver don't have several antennas. To meet the explosion of data transmission, multiple-input multiple-output (MIMO) technology has been widely used in wireless communication nowadays. MIMO system with multiple antennas at both the transmitter and receiver side brings benefits on data throughout and communication range with limited bandwidth and transmit power [67]. THz QKD with SISO and MIMO systems was introduced in Refs [61] and [68], with the main focus on mid-and far-infrared frequency ranges (10THz-40THz) at room temperature ($T=$ 296 K).

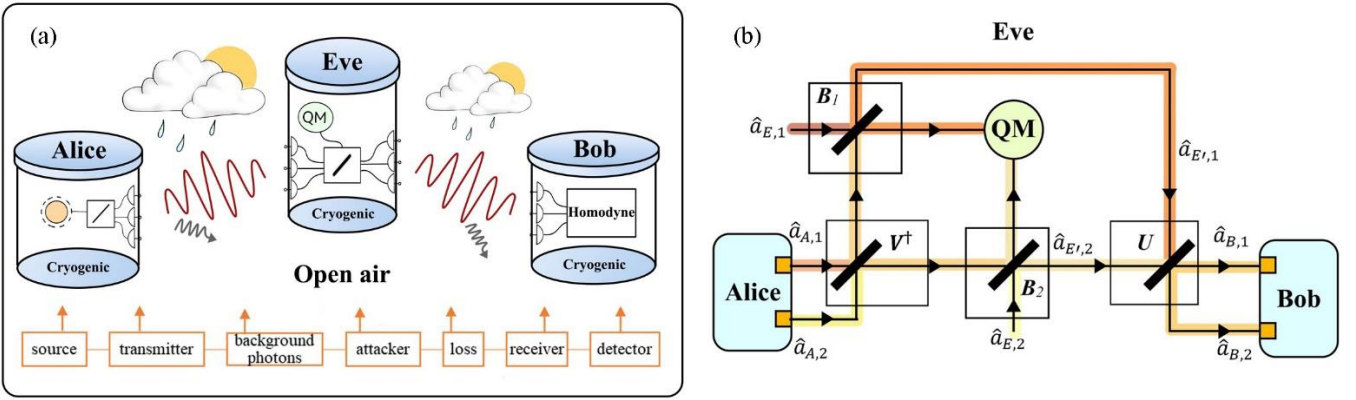

**Fig. 1.** (a) The system model of the proposed mm-waves and THz QKD. Alice prepares thermal states at the source which denotes a generator. The transmitter and receiver are antennas distributed in a one-dimensional uniform linear array (ULA). Alice, Bob, and Eve do their job at cryogenic ambient (low temperature). The channel loss through the open-air environment contains atmospheric absorption and free-space loss. Eve's output modes are stored in a quantum memory (QM). Bob uses a homodyne detector to measure quadrature. (b) The schematic of the phase shifters model described for a 2×2 MIMO system. The channel is modelled by 4 beam splitters. Alice generates 2 coherent states $\hat{a}_{A,1}$ and $\hat{a}_{A,2}$ based on random vectors and send them out from her 2 antennas as thermal states. These two states are mixed by the first beam-splitter $\boldsymbol{V}^{\dagger}$. Then Eve operates collective Gaussian attack. She prepares two two-mode squeezed vacuum states and uses beam-splitter $\boldsymbol{B}_1$ and $\boldsymbol{B}_2$ to combine the input and her states. The output and one of the original modes are saved in quantum memory (QM). Before Bob detects the input, the signals are mixed by beam-splitter $\boldsymbol{U}$. At last, Bob uses his two antennas to receive the modes $\hat{a}_{B,1}$ and $\hat{a}_{B,2}$.

Motivated by the works of [61] and [68], this work focuses on SISO and MIMO QKD at the temperature of $T<$ 50 K. We investigate CVQKD at frequency ranges of mm-waves and THz, from $f=$100 GHz to 1 THz, with antennas and detectors both operating in the cryogenic environment ($T<$ 50 K). Moreover, we compare the performance of both the SISO and MIMO CVQKD systems at this frequency range.

## II. SYSTEM MODEL

For the proposed mm-waves and THz quantum communication scheme, the cryogenic antennas generate electromagnetic (EM) fields that oscillate at an angular frequency ω. This EM field is quantized and the system gets a Hamiltonian $H = \hbar\omega(\hat{a}^{\dagger}\hat{a} + \frac{1}{2})$ . The $H$ is similar to the Hamiltonian of a quantum harmonic oscillator with $h$ as Planck's constant, $\hat{a}$ as the annihilation operator, and $\hat{a}^{\dagger}$ as the creation operator. Moreover, the quadrature field operators $\hat{q} = \frac{\hat{a}+\hat{a}^{\dagger}}{\sqrt{2}}$ and $\hat{p} = \frac{i(\hat{a}+\hat{a}^{\dagger})}{\sqrt{2}}$ are dimensionless canonical observables of the system (similar to the position and momentum of the quantum harmonic oscillator) [69]. Finally, coherent states of the system are the eigenstates of the annihilation operator $\hat{a}$, provided by $\hat{a}|\alpha\rangle = \alpha|\alpha\rangle$. Here, $\alpha = q + ip \in \mathbb{C}$ indicates the coherent state amplitude [68], taken from a two-dimensional Gaussian distribution. Two independent continuous variables $q$ and $p$ are used to create a secret key between Alice and Bob [61].

Notation: Boldface and italic capital letters such as $\boldsymbol{A}$ denote matrices. $\boldsymbol{A}^{\dagger}$ is the conjugate transpose of matrix $\boldsymbol{A}$ while $\boldsymbol{A}^{T}$ is the transpose. $\boldsymbol{0}_{M\times N} \in \mathbb{C}^{M\times N}$ is a zero complex matrix and $\boldsymbol{1}_{M\times N} \in \mathbb{C}^{M\times N}$ is a complex matrix of ones. $\boldsymbol{I}_M$ represents a $M \times M$ identity matrix. A $M \times M$ diagonal matrix described by $diag(\boldsymbol{a})$ with $\boldsymbol{a} \in \mathbb{C}^M$ shows $\boldsymbol{a}$ on its diagonal. And $\mathcal{N}(\mu, \boldsymbol{\Sigma})$ is a real multivariate Gaussian distribution in which the vector is $\mu$ and the covariance matrix (CM) is $\boldsymbol{\Sigma}$.

### A. Channel model

We consider a one-way communication channel to build a secret key between Alice and Bob as shown in Fig.1 (a). A MIMO mm-waves and THz communication channel between Alice and Bob include a transmitter with $N_t$ antennas at Alice's side and a receiver with $N_r$ antennas at Bob's side. We assume the antennas at both sides are distributed in a one-dimensional uniform linear array (ULA) with each antenna element's gain $G_a$. So the antenna gains of Alice and Bob are $G_t = N_tG_a$ and $G_r = N_rG_a$ [68]. The Gaussian modulation of the thermal state is a widely used encoding protocol for several frequencies [61]. Alice begins with a vacuum state $|0\rangle$ and generates $N_t$ coherent states $|a_i\rangle$ with amplitudes $a_i = Q_{A,i} + jP_{A,i}$, $i = 1,2,\dots,N_t$ from the $N_t$ antennas with quadratures being chosen from two independent random vectors $\boldsymbol{Q}, \boldsymbol{P} \sim \mathcal{N}(\boldsymbol{0}_{N_t\times 1}, V_s\boldsymbol{I}_{N_t})$ where $V_s$ is the variance of the initial signal encoding [68]. Two quadratures $\hat{Q}_{A,i}$ and $\hat{P}_{A,i}$ of a quantum THz source (thermal) state are randomly sent by the $i$-th antenna element of Alice and described by $\hat{X}_{A,i} \in \{\hat{Q}_{A,i}, \hat{P}_{A,i}\}$. So the $i$-th mode of Alice can be considered as $\hat{X}_{A,i} = a_i + \hat{0}$, where $\hat{0}$ is the thermal mode (quadrature operator) due to the background thermal noise at mm-waves and THz and $X_{Ai}$ denotes the classical modulated variable [61, 76]. The total variance of Alice's mode is

$$V_a = V_s + V_0 \tag{1}$$

where $V_0$ is the variance of thermal state (contains variance of vacuum mode and variance of preparation noise) [61]. $V_0$ is defined as

$$V_0 = 1 + 2\bar{n} \tag{2}$$

Here, 1 is the vacuum shot noise unit (SNU) and

$$\bar{n} = \left[\exp\left(\frac{hf_c}{k_B T}\right) - 1\right]^{-1} \tag{3}$$

is the mean thermal photon number, $h$ is Planck's constant, $k_B$ denotes Boltzmann's constant, $T$ is the environment temperature, and $f_c$ is the carrier frequency. Now, let's consider Alice sends her states to Bob (receiver) over an insecure quantum channel. Bob uses a noisy homodyne detection technique, which is based on mm-waves and THz shot-noise limited quantum detector that randomly switches between quadrature $\hat{Q}$ and $\hat{P}$, to measure the incoming thermal states.

The channel matrix between Alice and Bob could be modelled as [68, 70, 71]

$$\mathbf{H} = \sum_{l=1}^{L} \sqrt{\gamma_l} e^{j2\pi f_c \tau_l} \psi_{N_r}(\phi_{r,LOS}) \psi_{N_t}^{\dagger}(\phi_{t,LOS}) \tag{4}$$

where, $\mathbf{H} \in \mathbb{C}^{N_r \times N_t}$, $L$ is the full number of multipath components, $\tau_l$ is the propagation delay of the $l$-th multipath. We only consider the line-of-sight (LOS) path with $L=1$ in this work. So the path loss $\gamma_l$ is given by [68]

$$\gamma_{l=1} = G_t G_r \left(\frac{\lambda}{4\pi d}\right)^2 10^{-\frac{\delta d}{10}} \tag{5}$$

where $d$ is the distance (km) between Alice and Bob and $\delta$ is the atmospheric loss and is defined as dB/km [61, 70]. It contains both the free space path and the atmospheric absorption losses of mm-waves and THz waves. $\phi_{r,LOS}$ and $\phi_{t,LOS}$ are the angle of arrival seen by Bob, and the angle of departure from Alice, respectively. $\psi_K(\theta)$ represents the array response vector of a ULA which contains $K$ number of antennas.

The derivation details of the channel model are described by a singular-value decomposition (SVD) scheme introduced by Ref. [68] are presented in Appendix A.

Although the coherent attack is the general attack, the works reported in Refs. [72, 73] proved that once the system is secure against collective attacks, it is also secure against general attacks with the long secret key. In CVQKD, the most realistic and studied collective attack against Gaussian protocols is the entangling cloner attack [61]. So, we assume the channel are totally under Eve's control and she uses entangling cloners to steal information. Fig.1 (b) shows a $2 \times 2$ MIMO system built by 4 beam-splitters as an example [69]. After the two transmitted modes from Alice are combined by beam-splitter $\mathbf{V}^{\dagger}$, Eve will pick up two produced output modes. Eve should prepare two pairs of entangled Einstein-Podolsky-Rosen $\{\hat{e}_1, \hat{E}_1\}$ and $\{\hat{e}_2, \hat{E}_2\}$ (known also as two-mode squeezed vacuum states) in advance. Once received the input, Eve uses $\mathbf{B}_1$ and $\mathbf{B}_2$ to combine them with $\hat{E}_1$ and $\hat{E}_2$. The relationship of input and output of $\mathbf{B}_i$ can be written as [68, 69]

$$\begin{bmatrix} \hat{a}_{out,1} \\ \hat{a}_{out,2} \end{bmatrix} = \begin{bmatrix} \sqrt{\eta_i} & \sqrt{1-\eta_i} \\ -\sqrt{1-\eta_i} & \sqrt{\eta_i} \end{bmatrix} \begin{bmatrix} \hat{a}_{in,1} \\ \hat{a}_{in,2} \end{bmatrix} \tag{6}$$

Here, $\eta_i$ is the round trip transmissivity of two port beam-splitter $\mathbf{B}_i$. Then Eve will save one of the outputs from every beam-splitters ($\hat{E}'_1, \hat{E}'_2$) and the original modes ($\hat{e}_1, \hat{e}_2$) in her quantum memory (QM) and measure the ancilla modes to exploit information when Alice and Bob completed their classical communication. The other two output modes will be combined by the beam-splitter $\mathbf{U}$ and sent to Bob. We assume Alice applies $\mathbf{V}$ as the base of beamforming at her end, and Bob employs $\mathbf{U}^{\dagger}$ as the base of decoding at his side. The whole model could be described by [68, 69]

$$\hat{a}_B = \mathbf{U}^{\dagger} \mathbf{H} \mathbf{V} \hat{a}_A + \mathbf{U}^{\dagger} \mathbf{U} \mathbf{S} \hat{a}_E \tag{7}$$

where $\hat{a}_B = [\hat{a}_{B,1}, ..., \hat{a}_{B,N_r}]^T$ is the received mode of Bob, $\hat{a}_A = [\hat{a}_{A,1}, ..., \hat{a}_{A,N_t}]^T$ is the transmitted mode of Alice, $\hat{a}_E = [\hat{a}_{E,1}, ..., \hat{a}_{E,N_t}]^T$ is the injected Gaussian mode of Eve.

$$\mathbf{S} = diag\{\sqrt{1-\eta_1}, ..., \sqrt{1-\eta_r}, \mathbf{1}_{(M-r)\times 1}\} \tag{8}$$

is a diagonal matrix with $M = \min(N_t, N_r)$. The calculation details could be found in Appendix A. The efficient channel between Alice and Bob can be disintegrated into $r$ parallel SISO channels by utilizing the SVD of $\mathbf{H}$ ($r$ is the rank of $\mathbf{H}$). In this case, the relation between channels' input and output can be written as [68]

$$\hat{a}_{B,i} = \sqrt{T_i}\hat{a}_{A,i} + \sqrt{1-T_i}\hat{a}_{E,i}, \qquad i = 1,2,...,r \tag{9}$$

Here, $T_i$ is the $i$-th non-zero eigenvalue of $\mathbf{H}^{\dagger}\mathbf{H}$ and could also be considered as the $i$-th transmissivity of the channel. Homodyne experiments on one of the randomly selected quadratures will be performed by Bob for every $r$-received mode and get the result detailed in Appendix A.

We set $W$ to the variance of the thermal noise introduced by Eve. So, Bob will receive the $i$-th mode with the shot-noise level $V(\hat{X}_{B,i}) = T_i V_0 + (1 - T_i)W$ [61]. $\hat{X}_{B,i}$ is Bob's received quadrature described by Eq. (22) in Appendix A. If Eve wants to completely hide in the background noise, she can use $W = V_0$. Then Bob will receive the shot-noise level $V(\hat{X}_{B,i}) = V_0$ which is the same as what Alice sent. Although the value of $V_0$ could be enlarged by frequencies below 1THz, we could gain $V_0 \approx 1\ SNU$ by cooling down the system to low temperatures.

### B. Secret key rate

In this mm-waves and THz CVQKD scheme, Alice and Bob replicate the previous quantum communication protocol several times to generate a string. Then they correct errors in their string by using reconciliation protocol. Direct reconciliation (DR) is the scheme in which Bob uses Alice's encoding string to prepare the key while reverse reconciliation (RR) is when Alice prepares the key based on Bob's decoding result. Alice and Bob use a reconciliation protocol to achieve privacy amplification to reduce the knowledge stole by Eve [68]. In this work, we use an asymptotic secret key rate to study the performance of RR protocol in mm-waves and THz QKD. In RR it is possible to achieve a positive secret key rate when the channel transmissivity is almost 0 while in DR it required a channel transmissivity larger than 0.5 [76], and is impractical in mm-waves and THz because of the high losses (both atmospheric absorption and path).

According to [68], the secret key rate of the MIMO system could be divided into $r$ SISO channels. So we first analyse the secret key rate of a SISO channel. In the RR scheme, Alice and Bob use the decoding outcomes from Bob's side to generate their secret key [74]. Alice and Bob could estimate the mutual information they shared, described as $I(a:b)$, and the accessible information of Eve, described as $I(E:b)$. The asymptotic secret key rate $R^{\rightharpoonup}$ could be described by the surplus information shared by Alice and Bob and is given by [62]

$$R^{\rightharpoonup} = I(a:b) - I(E:b) \tag{10}$$

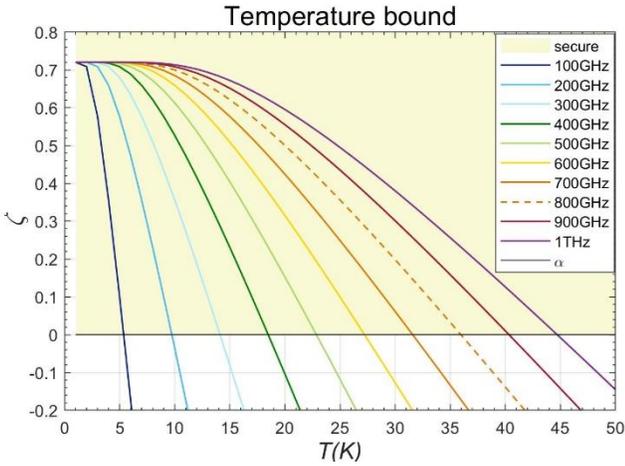

**Fig. 2.** The curves show $\zeta$ as a function of temperature for different frequencies. To achieve secure transmission, $\zeta$ should keep above line $\alpha$ (locate in the secure range).

The mutual information between Alice and Bob is

$$I(a:b) = \frac{1}{2} log_2 [1 + \frac{T V_s}{\Lambda(V_0, W)}] \qquad (11)$$

$$\Lambda(x,y) = Tx + (1-T)y \qquad (12)$$

while $V_s \gg V_0, W$, and Eve's information is bounded by the Holevo information $I(E:b)$ which is defined as

$$I(E:b) = H_E - H_{E|b} \qquad (13)$$

where $H_E$ and $H_{E|b}$ are the von Neumann entropy of Eve's total and conditional state given by Eq. (25) and Eq. (26) in Appendix A respectively [68, 75].

The total secret key rate is finally given by the sum of $r$ SISO links' secret key rate [68] assuming $T_i \to 0$

$$R_{MIMO}^{\rightarrow} = \sum_{i=1}^{r} R_i^{\rightarrow} \approx \zeta tr(\mathbf{H^\dagger H}) - rh(W) \qquad (14)$$

where

$$\zeta = 0.72 \left[ \frac{V_s}{W} - \ln\left(\frac{V_a + 1}{V_a - 1}\right)\left(\frac{V_a^2 - W^2}{2W} - V_a\right) \right] \qquad (15)$$

and $tr(\mathbf{H^\dagger H}) = \sum_{i=1}^{r} T_i$. $h(W)$ is a function given by

$$h(x) = \frac{x+1}{2} log_2\left(\frac{x+1}{2}\right) - \frac{x-1}{2} log_2\left(\frac{x-1}{2}\right) \qquad (16)$$

### C. System conditions

To achieve a positive secret key rate, $\zeta tr(\mathbf{H^\dagger H}) > rh(W)$ is required. Verified from Eq. (1)-(4), only $\zeta tr(\mathbf{H^\dagger H})$ in Eq. (14) depends on frequency and $\zeta$ also depends on temperature. With a given frequency, lower temperature could decrease $V_0$ and increase $\zeta$ in Eq. (15). This also happens with a given temperature and increasing frequency. But higher frequency normally brings higher path loss which may decrease $tr(\mathbf{H^\dagger H})$. Since $tr(\mathbf{H^\dagger H}) > 0$, the limit of $\zeta > \alpha = \frac{rh(W)}{tr(\mathbf{H^\dagger H})}$ is equal to $\zeta > \alpha = 0$ while $W=1$. This condition helps us find the balance between $\zeta$ and $tr(\mathbf{H^\dagger H})$ and get a positive secret key rate.

Figure 2 shows $\zeta$ as a function of temperature for different frequencies in a MIMO THz QKD system [68]. According to the simulation results, we achieved secure transmission for frequencies between $f$=100GHz and 1THz, at temperatures $T<$ 43K. The detail of the maximum operating temperature for each frequency is shown in Table 1.

### III. SIMULATION RESULT

We use the secret key rate ($R$) to define whether the system is safe or not. A positive secret key rate reflects a safe transmission. As $R$ of the MIMO system is the sum of the SISO system, we also compared the different performances of the MIMO and SISO systems at the same conditions.

TABLE I

| Frequency | Atmospheric absorption $\delta$ | Maximum temperature $T_{max}$ (K) | Distance of 32×32 MIMO system at $T_{max}$ (m) |
|---|---|---|---|
| 100GHz | 0.4dB/km | 4 | 700 |
| 200GHz | 3dB/km | 8 | 320 |
| 300GHz | 4dB/km | 13 | 75 |
| 400GHz | 20dB/km | 17 | 86 |
| 500GHz | 50dB/km | 21 | 68 |
| 600GHz | 150dB/km | 26 | 26 |
| 700GHz | 70dB/km | 30 | 38 |
| 800GHz | 100dB/km | 35 | 36 |
| 900GHz | 100dB/km | 39 | 25 |
| 1THz | 100dB/km | 43 | 21 |

### A. Simulation of the MIMO system

According to [60], [77], [78], the atmospheric absorption for each frequency is shown in Table 1. We assume the target of the secret key rate is $10^{-5}$ $bit/use$ [68]. Table 1 also shows the distance of 32×32 MIMO system at maximum temperature for each frequency. While $f$=700GHz and $f$=800GHz could reach more than 35m at $T$>30K, $f$=600GHz could only get to 26m at a lower temperature because of higher absorption at this frequency. Figure 3 shows the secret key rate as a function of secure transmission distance for frequencies from $f$=100GHz to 1THz at $T$=4K in a 32×32 MIMO system. The distance of $f$=100GHz and $f$=200GHz could achieve $d$=700m at $T$=4K which is the maximum through all frequencies. The secure distance for $f$=300GHz could get to $d$=500m but it will drop to under $d$=160m for frequencies above $f$=500GHz. If the number of antennas reduces to 8×8, the maximum distance could still reach $d$=200m for $f$=200GHz as shown in Fig. 4 (a). But for frequencies above $f$=600GHz in Fig. 4 (b), the distances are all below 50m because of the high absorption.

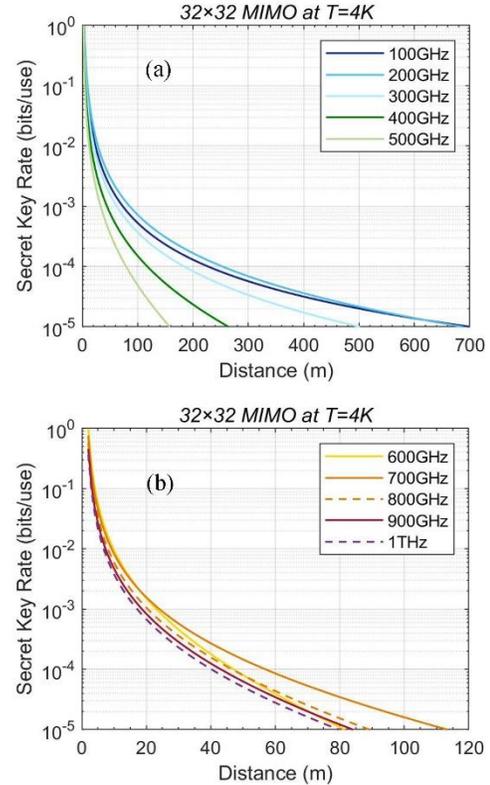

**Fig. 3.** (a) The transmission distance of a 32×32 MIMO system for $f$=100GHz to 500GHz at $T$=4K. (b) The transmission distance of a 32×32 MIMO system for $f$=600GHz to 1THz at $T$=4K. Here, parameters are $G_a$=30, $W$=1, $V_a$=1000.

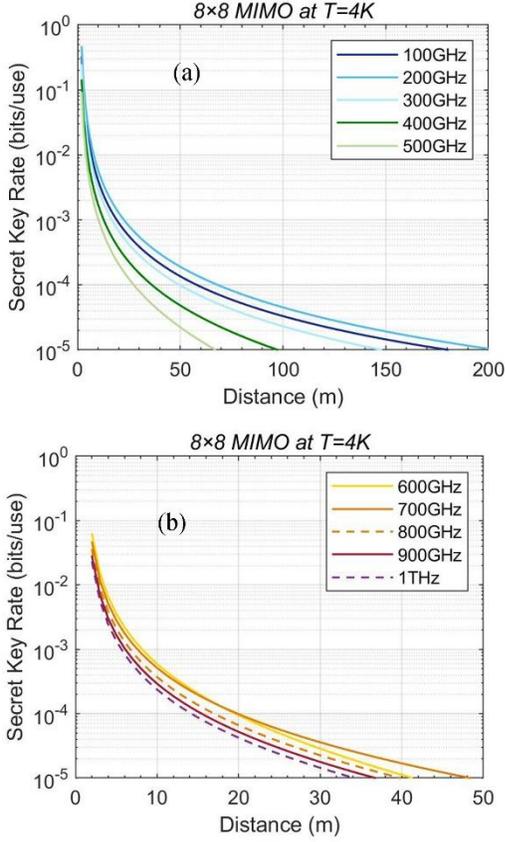

**Fig. 4.** (a) The transmission distance of an 8×8 MIMO system for $f$=100GHz to 500GHz under $T$=4K. (b) The transmission distance of an 8×8 MIMO system for $f$=600GHz to 1THz under $T$=4K. Here, parameters are $G_a$=30, $W$=1, $V_a$=1000.

High MIMO configuration could enhance the maximum secure distance as shown in Fig.5. While the maximum distance is $d$=700m for a 32×32 MIMO system, the secure transmission could achieve much more than $d$=8000m for a 1024×1024 MIMO at $T$=4K as shown in Fig. 5 (a). The same trend could be found for $f$=200GHz at $T$=8K and $f$=1THz at $T$=43K. For $f$=200GHz, the maximum distance is more than $d$=3000m for 1024×1024 MIMO at $T$=8K. But secure distances for $f$=1THz at $T$=43K are not as much shorter because of the high channel loss caused by atmospheric absorption and thermal noise. It could only get to 160m with a 1024×1024 MIMO antenna system at $T$=43K. And compared with Fig. 3 (b), the distance for $f$=1THz at $T$=43K is just one-fifth of $T$=4K, with 32×32 antennas.

### B. Simulation of the SISO system

Compared with a large MIMO scheme, the maximum distance of a SISO scheme is shorter at frequency ranges between $f$=100 GHz and $f$=1THz. We assume the target secret key rate is $10^{-5}$ $bit/use$ [68]. Figure 6 (a) shows that the maximum distance in the SISO system is less than 12m for $f$=200GHz at $T$=8K, which is much less than what is observed in a MIMO system shown in Fig. 5 (b). Figure 6 (b) shows the distance for $f$=100GHz-1THz in a SISO scheme at $T$= 4K. We could find that the maximum distance decrease with higher frequency at 200GHz-1THz range caused by the path loss. Comparing Figs. 6 (a) and (b), it is obverse that lower temperature increases the maximum distance for the same frequency.

Comparing Fig.6 (b) with Fig. 4 we find that at the same temperature, the MIMO scheme could achieve a longer distance than the SISO scheme. So it is necessary to use multiple antennas in mm-waves and THz QKD system. As frequencies $f$=800GHz, $f$=900GHz, and $f$=1THz have the same atmospheric absorption, Fig.6 (b) shows that the transmission distance improved by the lower frequency. This may cause by the increase of $tr(\boldsymbol{H}^{\dagger}\boldsymbol{H})$ as lower frequencies have less free-space path loss and matter more than the increase of $V_0$ (which in turn decreases $\zeta$) at this range. As a result, to enhance the secure distance, we could use more antennas, cool down the temperature or choose lower frequencies.

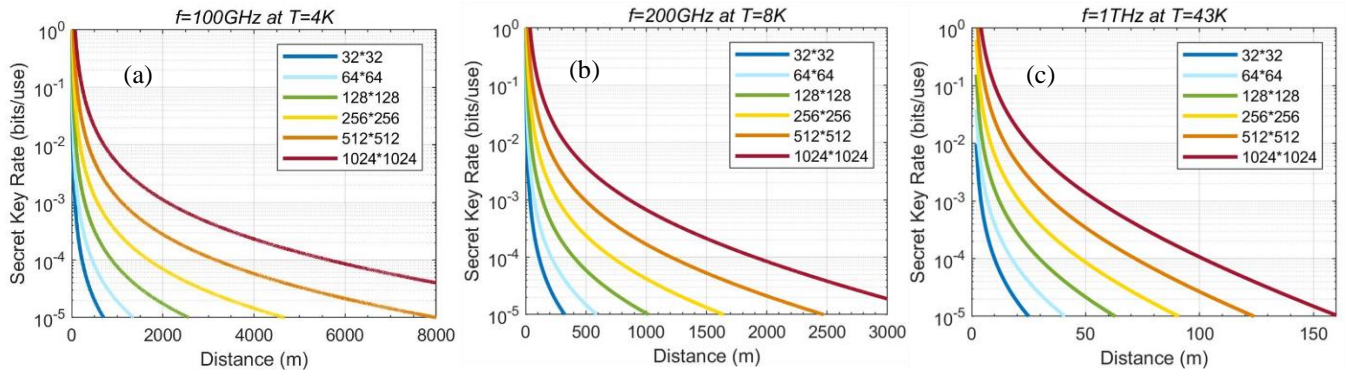

**Fig. 5.** The curves show the secret key rate as a function of transmission distance for the different numbers of antennas in MIMO systems. (a) is the result for $f$=100GHz, at $T$=4K. (b) is the result for $f$=200GHz, at $T$=8K. and (c) is the result for $f$=1THz, at $T$=43K.

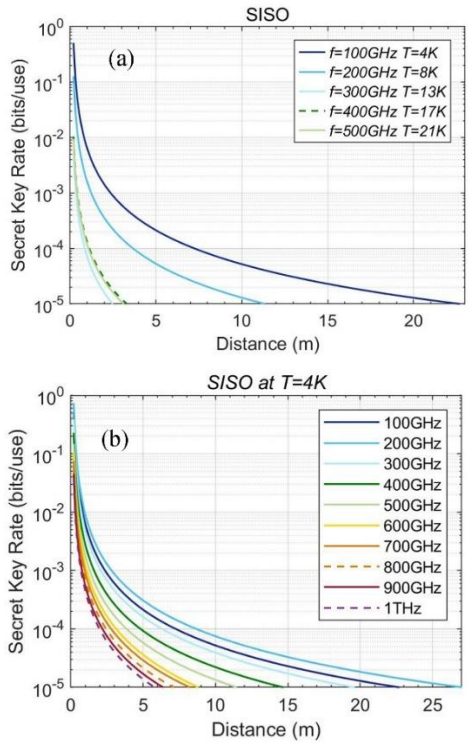

**Fig. 6.** (a) The secret key rate R as a function of transmission distance for SISO for frequency ranges between $f$=100GHz and $f$=500GHz at the highest operational temperature. Atmospheric absorptions in the channel are shown in table 1. (b) The secret key rate R as a function of transmission distance for SISO for frequency ranges between $f$=100GHz and $f$=1THz at $T$=4K.

## IV. CONCLUSION

In this work, we studied the SISO, and MIMO QKD schemes in mm-waves and THz frequency ranges, from $f$= 0.1 to $f$= 1 THz, with the cryogenic operation of sources and detectors. We found that a positive secret key rate can be observed at the targeted frequency range. MIMO technology could improve the secure transmission distance compared with the SISO scheme. Moreover, we showed that the more antennas are in the system the longer transmission distance could be achieved. To build a long-way secure communication channel, more antennas, low-temperature operation and lower frequencies are required. We assumed the perfect quality of the beam-splitter and communication channel in this study, however, there are still many research efforts to be performed to accomplish the whole scheme in practical hardware.

# APPENDIX A

## A. Computation of channel model

The channel matrix between Alice and Bob could be modelled as Eq. (4)

$$H = \sum_{l=1}^{L} \sqrt{\gamma_l} e^{j2\pi f_c \tau_l} \psi_{N_r}(\phi_{r,LOS}) \psi_{N_t}^\dagger(\phi_{t,LOS})$$

where, $H \in \mathbb{C}^{N_r \times N_t}$ , $L$ is the full number of multipath components, $\tau_l$ is the propagation delay of the $l$-th multipath. The path loss $\gamma_l$ for the line-of-sight (LOS) path ($L=1$) is given by Eq. (5)

$$\gamma_{l=1} = G_t G_r \left(\frac{\lambda}{4\pi d}\right)^2 10^{-\frac{\delta d}{10}}$$

where $d$ is the distance (km) between Alice and Bob and $\delta$ is the atmospheric loss. $\phi_{r,LOS}$ and $\phi_{t,LOS}$ is the angle of arrival seen by Bob, and the angle of departure from Alice, respectively. $\psi_K(\theta)$ represents the array response vector of a ULA which contains $K$ number of antennas [68]:

$$\psi_K(\theta) = \frac{1}{\sqrt{K}} \left[ 1, e^{j2\pi \frac{d_a}{\lambda} sin\theta}, \ldots, e^{j2\pi \frac{d_a}{\lambda}(K-1)sin\theta} \right]^T \quad (17)$$

Here, $d_a$ and $\lambda$ denote the inter-antenna spacing and wavelength of the carrier signal, respectively. We assume $d_a = \frac{\lambda}{4}$ in this work. The channel model could be described by a singular-value decomposition (SVD) scheme introduced by Ref. [68]:

$$H = U\Sigma V^\dagger \quad (18)$$

Here, $U \in \mathbb{C}^{N_r \times N_r}$ and $V \in \mathbb{C}^{N_t \times N_t}$ are unitary matrices represent $K$ phase shifters [69], and $\Sigma$ is given as

$$\Sigma = \begin{bmatrix} diag\{\sqrt{\eta_1}, \ldots, \sqrt{\eta_r}\} & \mathbf{0}_{r \times (N_t - r)} \\ \mathbf{0}_{(N_r - r) \times r} & \mathbf{0}_{(N_r - r) \times (N_t - r)} \end{bmatrix} \quad (19)$$

where $r$ is the rank of $H$ and $\sqrt{\eta_1}, \ldots, \sqrt{\eta_r}$ are the $r$ non-zero singular values of $H$ generated by the transmissivity of beam-splitters in the system [69].

As shown in Fig.1 (b), the whole model of a MIMO system could be described by [68, 69]

$$\hat{a}_B = U^\dagger H V \hat{a}_A + U^\dagger U S \hat{a}_E \quad (20)$$

where $\hat{a}_B = [\hat{a}_{B,1}, \ldots, \hat{a}_{B,N_r}]^T$ is the received mode of Bob, $\hat{a}_A = [\hat{a}_{A,1}, \ldots, \hat{a}_{A,N_t}]^T$ is the transmitted mode of Alice, $\hat{a}_E = [\hat{a}_{E,1}, \ldots, \hat{a}_{E,N_t}]^T$ is the injected Gaussian mode of Eve.

$$S = diag\{\sqrt{1 - \eta_1}, \ldots, \sqrt{1 - \eta_r}, \mathbf{1}_{(M-r) \times 1}\} \quad (21)$$

is a diagonal matrix with $M = \min(N_t, N_r)$. We get $U^\dagger U = I_{N_r}$ and $V^\dagger V = I_{N_t}$ as $U$ and $V$ are unitary matrices. Considering Eq. (18) $H = U\Sigma V^\dagger$, diagonal matrix $\Sigma$ and $S$, Eq. (20) would turn to $\hat{a}_B = U^\dagger U \Sigma V^\dagger V \hat{a}_A + U^\dagger U S \hat{a}_E = \Sigma \hat{a}_A + S \hat{a}_E$ . So the efficient channel between Alice and Bob can be disintegrated into $r$ parallel SISO channels.

Bob randomly chooses quadratures for his input and operates homodyne measurements. Then the input-output relationship of $i$-th parallel channel between Bob's received quadrature $\hat{X}_{B,i}$, and Alice's transmitted quadrature $\hat{X}_{A,i}$, can be written by a generic quantum channel as [68]

$$\hat{X}_{B,i} = \sqrt{T_i} \hat{X}_{A,i} + \sqrt{1 - T_i} \hat{X}_{E,i}, \qquad i = 1, 2, \ldots, r \quad (22)$$

with $T_i$ and $\hat{X}_{E,i}$ as transmittance, and Eve's excess noise quadrature, respectively. We can write for Eve's ancilla mode:

$$\hat{X}_{E',i} = -\sqrt{1 - T_i} \hat{X}_{A,i} + \sqrt{T_i} \hat{X}_{E,i}, \qquad i = 1, 2, \ldots, r \quad (23)$$

## B. Computation of the secret key rate

The secret key rate $R^\blacktriangleleft = I(a:b) - I(E:b)$ is given defined by mutual information. Assuming Gaussian statistic for simulating purposes, the mutual information between Alice and Bob is

$$I(a:b) = \frac{1}{2} log_2[1 + \frac{TV_s}{\Lambda(V_0, W)}]$$

$$\Lambda(x, y) = Tx + (1 - T)y$$

And Eve's information is bounded by the Holevo information $I(E:b)$ which is defined as

$$\mathrm{I(E:b)} = H_E - H_{E|b} \tag{24}$$

$$H_E = h(v_1) + h(v_2) \tag{25}$$

$$H_{E|b} = h(v_3) + h(v_4) \tag{26}$$

where $H_E$ and $H_{E|b}$ are the von Neumann entropy of Eve's total and conditional state respectively [68, 75]. The von Neumann entropy depends on symplectic eigenvalues is given by [68]

$$v_1 = \Lambda(W, V_a), v_2 = W, \tag{27}$$

$$v_3, v_4 = \sqrt{\frac{1}{2}\left(\Delta \pm \sqrt{\Delta^2 - 4\Upsilon}\right)}, \tag{28}$$

and

$$\Delta = \frac{V_a W \Lambda(W, V_a) + W \Lambda(W V_a, 1)}{\Lambda(V_a, W)} \tag{29}$$

$$\Upsilon = \frac{V_a W^2 \Lambda(W, V_a) \Lambda(W V_a, 1)}{\Lambda^2(V_a, W)} \tag{30}$$

The function h(x) is defined as

$$\mathrm{h(x)} = \frac{x+1}{2} log_2 \left(\frac{x+1}{2}\right) - \frac{x-1}{2} log_2 \left(\frac{x-1}{2}\right) \tag{31}$$